\documentclass[pra,aps,twocolumn,showpacs,,amsmath,amssymb]{revtex4-1}
\usepackage{amsmath,graphicx}
\usepackage{bm}
\usepackage{amssymb}
\usepackage{graphicx,xcolor}
\usepackage{epigraph}
\usepackage{csquotes}

\usepackage{amsmath,amssymb}
\begin{document}
\def\e{\enquote}
\def\tr{\rm{Tr}}
\def\la{{\langle}}
\def\ra{{\rangle}}
\def\a{{\alpha}}
\def\q{\quad}
\def\w{\tilde}
\def\om{\omega_L}
\def\t{\tilde{t}}
\def\a{\hat{A}}
\def\h{\hat{H}}
\def\E{\mathcal{E}}\def\la{{\langle}}
\def\u{\hat U}
\def\tt{y}
\def\U{\hat U}
\def\C{\hat C}
\def\D{\Delta}
\def\S{\tilde S}
\def\A{\mathcal{A}}
\def\B{{\textbf B}}
\def\QQ{\hat S}
\def\R{\text {Re}}
\def\e{\enquote}
\def\qq{s}
\def\Q{S}
\def\fb{\overline F}
\def\wb{\overline W}
\def\nl{\newline}
\def\h{\hat H}
\def\ff{\overline q}
\def\k{\overline k}
\def\F {Q}
\def\f{q}
\def\r{\color{black}}
\def\lm{\lambda}
\def\lmu{\underline\lambda}
\def\q{\quad}
\def\t{\tau}
\def\l{\ell}
\def\n{\\ \nonumber}
\def\ra{{\rangle}}
\def\Ep{{\mathcal{E}}}
\def\T{{\mathcal{T}}}
\def\M{{\mathcal{M}}}
\def\omga{{\epsilon}}
\def\t{{\tau}}
\def\h{\hat{H}}
\title{Wave packets,  \e{negative times} and the elephant in the room}
%\author {D. Sokolovski} \author {E. Akhmatskaya}

\author {D. Sokolovski$^{1,2}$ (corresponding author)} 
\author{A. Matzkin$^{3}$}
%\email {dgsokol15@gmail.com}
%\author {E. Akhmatskaya$^{2,3}$ } 
\affiliation{$^1$ Departmento de Qu\'imica-F\'isica, Universidad del Pa\' is Vasco, UPV/EHU, Leioa, Spain}
%\affiliation {$^2$ Basque Center for Applied Mathematics (BCAM),\\ Alameda de Mazarredo 14, 48009 Bilbao, Bizkaia, Spain}
\affiliation{$^2$ IKERBASQUE, Basque Foundation for Science, Plaza Euskadi 5, 48009 Bilbao, Spain}\date{\today}
\affiliation{$^{3}$ Laboratoire de Physique Th\'eorique et Mod\'elisation, CNRS Unit\'e
8089, CY Cergy Paris Universit\'e, 95302 Cergy-Pontoise cedex, France}
\email{dgsokol15@gmail.com}
%\affiliation{$^c$ 
%Department of Particle Physics and Astrophysics, Weizmann Institute of Science, Rehovot, 76100, Israel}
\begin{abstract}
\noindent
Controversy surrounding the \e{tunnelling time problem} stems from the seeming inability of quantum mechanics to provide, in the usual way, a definition of the duration a particle is supposed
to spend in a given region of space. For this reason, the problem is often approached from an \e{operational} angle.  One such approach  uses the position of the transmitted wave packet in order to 
infer the duration the particle spends in the barrier. Here we replace the barrier with a tuneable Mach-Zehnder interferometer (MZI). With this analogy one is able, at least in principle, to achieve any advance
or delay of the wave packet sent to the chosen outgoing port. The Uncertainty Principle prevents one from combining the durations spent in each arm of the MZI into a meaningful duration when 
both arms are engaged. There is no justification for invoking \e{superluminal} or \e{negative} times, since the particle is able to arrive at the same position (and with a higher probability)
if the same initial state propagates through only one arm of the MZI. The same is true in the case of tunnelling,
 where the transmitted wave packet results from destructive interference between delayed  multiple copies of the free state.
 % delayed relative to the free propagation.
\end{abstract}
%\pacs{03.65.Ta, 03.65.AA, 03.65.UD}
%{Foundations of quantum mechanics}
%\pacs{03.65.AA}{Quantum systems with finite Hilbert space}
%\pacs{03.65.UD}{Entanglement and quantum nonlocality}
\maketitle

\section{Introduction}
It is a commonplace to state that the tunnelling time problem, i.e., the problem of determining the duration $\t$ a tunnelling particle 
spends in the barrier, continues to cause controversies as of now.
% (for reviews see, e.g., \cite{Rev1}-\cite{Rev7}.
In standard quantum mechanics, there have been at least three main approaches, all trying to extend classical reasoning to an essentially 
quantum context. In 1932 McColl proposed deducing the duration from the position of the transmitted wave packet \cite{WP0}.
The method was later extended in  \cite{WP1}-\cite{WP2}. Thirty five years later Baz' \cite{L1} proposed equipping the scattered particle 
with a magnetic moment (spin), and determining $\t$ by dividing its angle of rotation in a small magnetic field by the Larmor frequency $\omega_L$. 
Finally, in 1982 B\"uttiker and Landauer suggested extracting the duration spent in the barrier from the adiabatic limit reached in a small oscillatory 
field added to the barrier \cite{BL}.
The literature on the subject is extensive, and we refer the reader to various reviews
(see, e.g., \cite{REV1}-\cite{REV7}). 
 None of the three methods mentioned so far are free from problems
(for a recent discussion of the B\"uttiker-Landauer time see \cite{DSBL})
Below we will be interested mostly in the wave packet technique \cite{WP0}-\cite{WP2}, to which we will return after a brief digression. 
%%%%%%%%%%%%%%%%%%%%%%%%%%%
\section{The elephant in the room}
Perhaps the most straightforward analysis of the problem can be given in the case of the Larmor (Baz') clock \cite{DSL1}.
Feynman paths connecting the particle's initial and final states can be grouped
into \e{pathways} according to the duration 
$\t$ a path spends on the region $\Omega$ where the magnetic field is introduced. 
By summing the corresponding path amplitudes one obtains the amplitude $\A(\t)$ of spending in $\Omega$
$\tau$ seconds, The spin's final state is, therefore, a superposition of the copies of its initial state, rotated by by all $\omega_L\t$.
\newline
One basic difficulty is now apparent. Tunnelling is an interference phenomenon.  The small tunnelling amplitude is a result
of destructive interference between the pathways. The Uncertainty Principle \cite{FeynL} - the elephant in the room where a tunnelling time discussion is held- 
 forbids knowing the path taken by the particle without destroying the interference pattern, i.e., by destroying the very tunnelling event one wishes 
 to study. 
 \newline
 The difficulty  is already evident if the number of pathways reduced to two \cite{DSL2}. Consider  a Mach-Zehnder interferometer (MZI), 
with a small magnetic field introduced in such a way that a particle with a spin spends in it durations $\t_1$ and $\t_2$, if travelling via 
its first and the second arm, respectively. On exit from the interferometer, at a location which the particle can reach via each arm with 
amplitudes $\A_1$ and $\A_2$,  the spin's angle of rotation is given by \cite{DSL2}
\begin{eqnarray}\label{1}
\varphi =\om\R\left [\frac{\t_1\A_1+\t_2\A_2}{\A_1+\A_2}\right ].
\end{eqnarray}
With only a few  {\it apriori} restrictions on the amplitudes $\A_{1,2}$, the real part of the 
\e{complex time} in the square brackets is not an acceptable duration. It can be zero, prompting suspicion 
of violation of Einstein's relativity, or negative, or even exceed the time it takes the particle to travel to the chosen location \cite{DSL2}.
Whoever wishes to avoid unsavoury sensationalism, is forced to accept impossibility of inferring a meaningful duration 
by combining $\t_1$ and $\t_2$, unless the interference between the paths is destroyed.
%The sensationalist can declare these bizarre conclusions to be the features of the brave new quantum world. 
%A more rational person is forced to admit to a failure of combining  durations $\t_1$ and $\t_2$ into something meaningful, unless the interference is destroyed. 
%
This must be, we note, a consequence of the Uncertainty Principle \cite{FeynL}. 
\newline
Below we will show that the same problem will occur if, instead of employing Larmor precession, one chooses to
monitor the position of the wave packet exiting the interferometer. As with the Larmor clock, the  Uncertainty Principle is responsible - the same elephant in a different room. 
%%%%%%%%%%%%%%%
\section{Wave packet delays}
Consider a Gaussian wave packet (WP)
\begin{eqnarray}\label{1a}
 G(x) =(\pi \Delta x^2/2)^{-1/4}\exp(-x^2/\Delta x^2)
 \end{eqnarray}
 injected into the Mach-Zehnder interferometer   shown in Fig.1a.
 %%%%%%%%%%%%%%%%%%%
\begin{figure}[h]
\includegraphics[angle=0,width=8.5cm, height= 5cm]{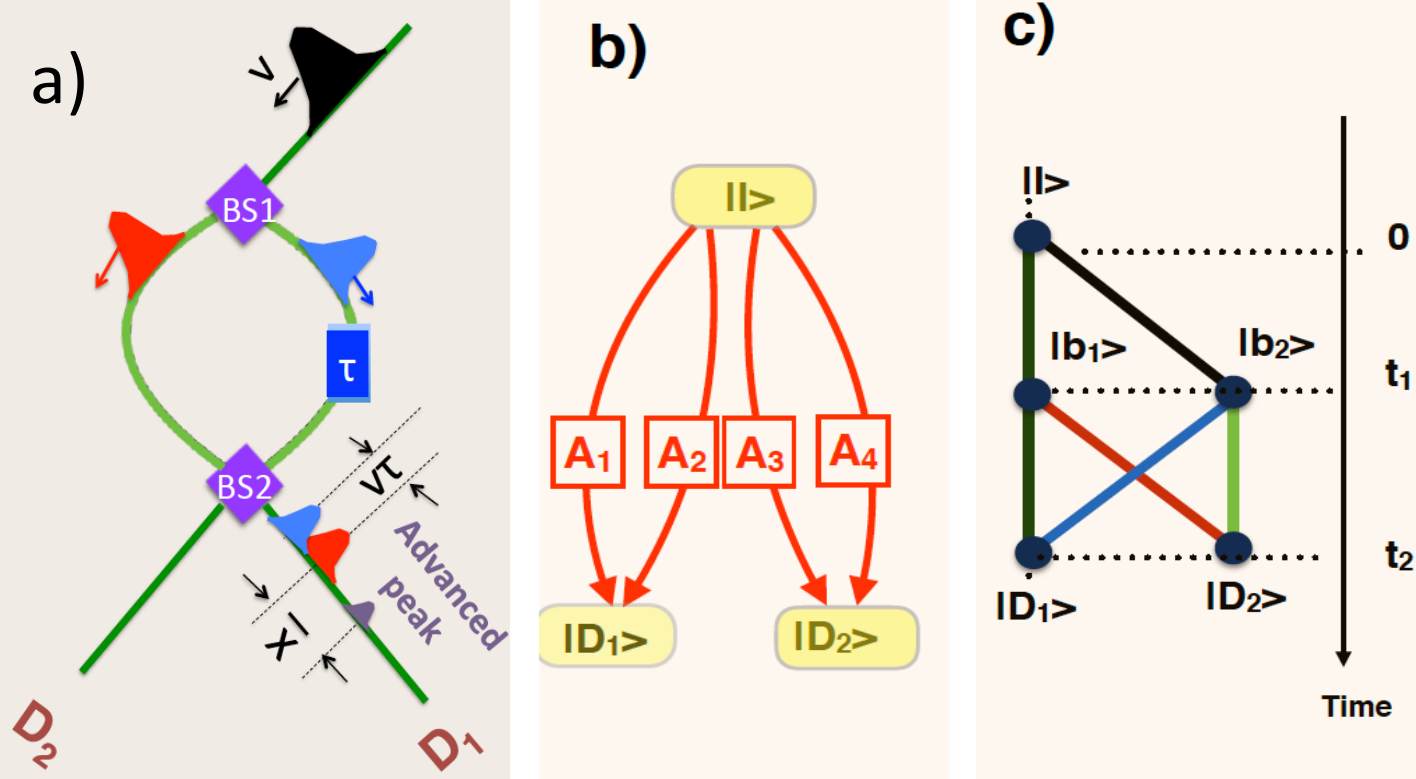}
\caption {a) A wave packet is divided into two parts which are recombined again while passing through a pair 
of beam splitters (BS). A delay by $\t$ sec. occurs for the part passing through  the right route. 
A wave packet travelling to detector $D_1$ consists of two parts (cf. Eq.(\ref{2a}) which may or may not overlap, depending on $v$, $\t$, 
and the initial wave packet's width $\Delta x$. Their sum can result in a much smaller advanced Gaussian. 
b) A path diagram of the system in a) and the four amplitudes $\A_i$. The paths leading to different final states $D_1$ and $D_2$ never interfere.
c) An equivalent quantum measurement problem (see Sect.IV). }
\end{figure}
%%%%%%%%%%%%%%%%
\newline
%At $t=0$ particle be in a wave packet state $|G\ra$.
The WP, propagating at velocity $v$, splits at the first beamsplitter ($BS$), experiences a delay by $\tau$ secs in the 
MZI's right  arm, its parts recombined after the second $BS$.
 %and leave it after the second BS, with part of it moving  towards detector $D_1$.
For simplicity, we neglect WP's spreading and set the new origin, $x=0$, at the centre of 
 the WP, propagating along the left  arm, (see schematic in Fig.1a). 
%t  atime $T$, t
The part of the wave function travelling towards detector $D_1$, is given by
\begin{eqnarray}\label{2a}
G_1
%_{D_1}
(x) = 
%A_1 G_0(x,T) + A_2 G_\t(x,T)\equiv\n
 \A_1 G(x) + \A_2 G(x+v\t),
\end{eqnarray}
where 
%$G(x,T)=\la x|G(t)\ra$ is the result of propagating $|G(0)\ra$ along the first arm (path), and
 $\A_1$ and $\A_2$ are the path amplitudes, determined by the properties of the beamsplitters, etc. 
%\newline
The width of the initial state now controls the amount of interference 
 between the two terms
in (\ref{2a}).
\newline
For  $\Delta x << v\t$  the Gaussians in  (\ref{2a}) do not overlap, the probability to find the 
particle at $x$, $P(x)=|\A_1|^2|G(x)|^2+|\A_2|^2|G(x+v\t)|^2$ is a sum of two disjoint terms, 
so one always knows the path travelled by the particle found at  a particular $x$. Interference between the pathways is destroyed
and, in accordance with the Uncertainty Principle, the probability to reach the detector $D_1$ is given by 
%\begin{eqnarray}\label{5}
$P=\int dx P(x) \approx |A_1|^2 + |A_2|^2$.
%\end{eqnarray}
%\newline

In the opposite limit $\Delta x >> v\t$ one finds
$G(x+v\t)\approx G(x)$, interference between the pathways is duly restored and 
in, accordance with the Uncertainty Principle, the probability to reach $D_1$ is 
$P \approx |A_1 + A_2|^2$. There is not a single $x$, such that the particle found there 
can be said to travel one path, and not the other. Moreover, it is a well known property of Gaussians \cite{Vaid} that 
as $\Delta x \to \infty$,  
\begin{eqnarray}\label{3a}
P(x)
%\to
 %
  \xrightarrow[\Delta x \to 0] {}
   |\A_1 + \A_2|^2\exp\left [ -2\frac{(x-\overline x)^2}{\Delta x^2}\right],
 \end{eqnarray}
 with the positions of the peak (and the centre of mass) of the Gaussian in the r.h.s. of (\ref{3a})  given by an expression somewhat similar to (\ref{1}) 
 \begin{eqnarray}\label{4a}
\overline x \equiv-\R\left [\frac{v\t\A_2}{\A_1 + \A_2}\right].
 \end{eqnarray}
 Convergence of $P(x)$ to the single-peak form (\ref{3a}), described by the Catastrophe Theory \cite{CAT}, can be shown to be point-wise \cite{DSA}.
 \newline
Next we note that one can, at least in principle, choose the delay $\tau$ and the amplitudes $\A_{1,2}$ so as to give
the distance between the peak propagating via both arms,  and the peak travelling via the left arm only, 
 $\overline x$,  any desired value. 

%%%%%%%%%%%%%%%%
\section{Any delay
%$\t_{delay}$
 is possible}
The system shown in Figs.1a and 1b is described by the total of four paths, two for each final state $D_1$ and $D_2$. The four complex valued path amplitudes, $\A_i$, $i=1,2,3,4$ 
need to satisfy the probability conservation rules, {\r whether or not interference between is destroyed,} 
 \begin{eqnarray}\label{b1}
\sum_{i=1}^4|\A_i|^2=|\A_1 + \A_2|^2+|\A_3 + \A_4|^2=1,
\end{eqnarray}
However,  little restriction is imposed on the signs of their real and imaginary parts.  Thus, for a given real $y$, 
one can always find $\A_i$ such that 
 \begin{eqnarray}\label{b2}
\R\left [\frac{y\A_1}{\A_1 + \A_2}\right]=z
 \end{eqnarray}
for any desired  $z$.  Indeed,  Eq.(\ref{2a}) also describes the state of a Gaussian von Neumann pointer \cite{vN}, set up to measure a projector, 
$\hat B= y |b_2\ra\la b_2|$, for a two-level system (qubit), pre- and post-selected in some states $|I\ra$ and $|D_1\ra$, respectively \cite{DSS}.
In this case, the amplitudes of the four paths shown in Figs. 1b and 1c, are known to be given by \cite{DSS} (neither the pointer, nor the qubit,  have own dynamics)
 \begin{eqnarray}\label{b3}
\A_1=\la D_1|b_1\ra\la b_1|I\ra, \n
\A_2=\la D_1|b_2\ra\la b_2|I\ra, \n
\A_3=\la D_2|b_1\ra\la b_1|I\ra, \n
\A_4=\la D_2|b_2\ra\la b_2|I\ra.  
\end{eqnarray}
 {\r  Constructed in this manner, the amplitudes (\ref{b3}) automatically satisfy the conditions (\ref{b1}). One only needs to show that,
for a given pair of final states $|D_{1,2}\ra$, and a basis $|b_i\ra$, it is always possible to find a suitable pair of orthonormal final states $|I\ra$.
There are various choices,
 including the one which will keep $\A_i$ real.  e.g., 
 \begin{eqnarray}\label{b4}
\la b_1|D_1\ra = N^{-1}/\la b_1|I\ra^*,\q\q\q\q\q\q\q \n
\la b_2|D_1\ra =N^{-1}\left (\frac{z}{\tt-z}\right )/\la b_2|I\ra^*,\q\q\q \n
\la b_1|D_2\ra = \la b_2|D_1\ra^*,\q\q\q\q\q\q\q\q\q\n
%\q\q\q\q  \n
\la b_2|D_2\ra =-\la b_1|D_1\ra^*,\q\q\q\q\q\q\q\q
% \sqrt{\frac{1}{|\la b_1|I\ra|^{2}} +\frac{}{(\tt-z)^2 |\la b_2|I\ra|^{2}}}.\q\q
\end{eqnarray} 
where $N=|\la b_1|I\ra|^{-1}(\tt-z)^{-1}\sqrt{(\tt-z)^2 + z^2}$.
%q\q\q\q\q\q\q\q\q
In the special case $|D_{1,2}\ra=(|b_1\ra\pm |b_2\ra)/\sqrt 2$  the four amplitudes
in (\ref{b3}) become
 \begin{eqnarray}\label{b5}
%\A_1=\frac{1}{\sqrt{2[1+\frac{z^2}{(\tt-z)^2 }]}}, \q
%\q\q\q\q\q \n
\A_1=\frac{(y-z)}{\sqrt{2[(\tt-z)^2+z^2] }}, \q
\q\q\q\q\q \n
\A_2= \frac{z}{y-z} A_1,\q\q\q\q\q\q\q\q\q\q\n
\A_3=A_1, \q
\A_4=-A_2.\q\q\q\q\q\q\q
\end{eqnarray} 
We then note that for $z=0$. $A_2=A_4=0$, and the system arrives in $|D_1\ra$ and $|D_2\ra$ via $|b_1\ra$ (the first and the third paths in Fig.1b).
% and in $|D_2\ra$ via $|b_2\ra$ (the fourth path in Fig.1b).
Similarly, for $z=y$, both final states $|D_{1,2}\ra$ are reached via $|b_{2}\ra$ (the second and the fourth paths in Fig.1b).}
For a particle in Fig.1a we put $y=-v\tau$, and, wishing only to illustrate a principle, assume that the amplitudes in Eqs (\ref{b3}) can somehow be realised in practice.
% and $|b_{1}\ra$, respectively.
%\newline
%Finally, as  $z\to\t$, $A_1\to -A_2\to 1/2$ and so that detection in $|D_1\ra$ becomes very unlikely and, since $A_3\to A_4 \to 1/2$, the probability of detection in 
%$|D_2\ra$ tends to $|A_3 + A_4|^2 \to 1$. 
%%%%%%%%%%%%%%%%%%%%%%%%%%%%%
%\section{The foolishness: zero and  negative times}
%\section{How to turn a delay into speed up (by reshaping)}
\section{Advancing the peak}
Now the following experiment is, at least in principle, possible.
First one removes the beamsplitters in Fig.1a, 
%replaces the $BS2$ with a perfect mirror, 
and
lets a broad  Gaussian WP travel towards $D_1$ along the left arm of the MZI until time $T$, and determines the position of its peak (or of the centre of mass). 
Next the beamsplitters are re-introduced, so a part of the initial WP is diverted in through the right arm, where it is delayed by $\t$, and then recombined with the part travelling the left arm.
The result is the Gaussian in Eq.(\ref{3a}),  whose peak's position $\overline x$ is then compared to that of the WP travelling along the left arm only, $|G(x)|^2$. 
As was demonstrated above, with an appropriate choice of the amplitudes $\A_{1.2}$ the peak can be made to lie {\it ahead} of the free one, $\overline x > 0$, (or arrive earlier at a fixed detector, if one so wishes). To observe this \e{speed up} produced by the {\it delay} in the right arm, one does not even need to make initial WP very broad. 
Transformation of the twin-peak density $P(x)$ into a single-peak form as the width $\Delta x$ increases is followed in Fig.2. 
%It corresponds to cusp singularity of the Catastrophe Theory \cite{DS2}, and
 A reasonable approximation to Eq.(\ref{3a}) is obtained already for $\Delta x/v \t =5$, as is shown in Fig.3.
  \begin{figure}[h]
\includegraphics[angle=0,width=8.5cm, height= 6cm]{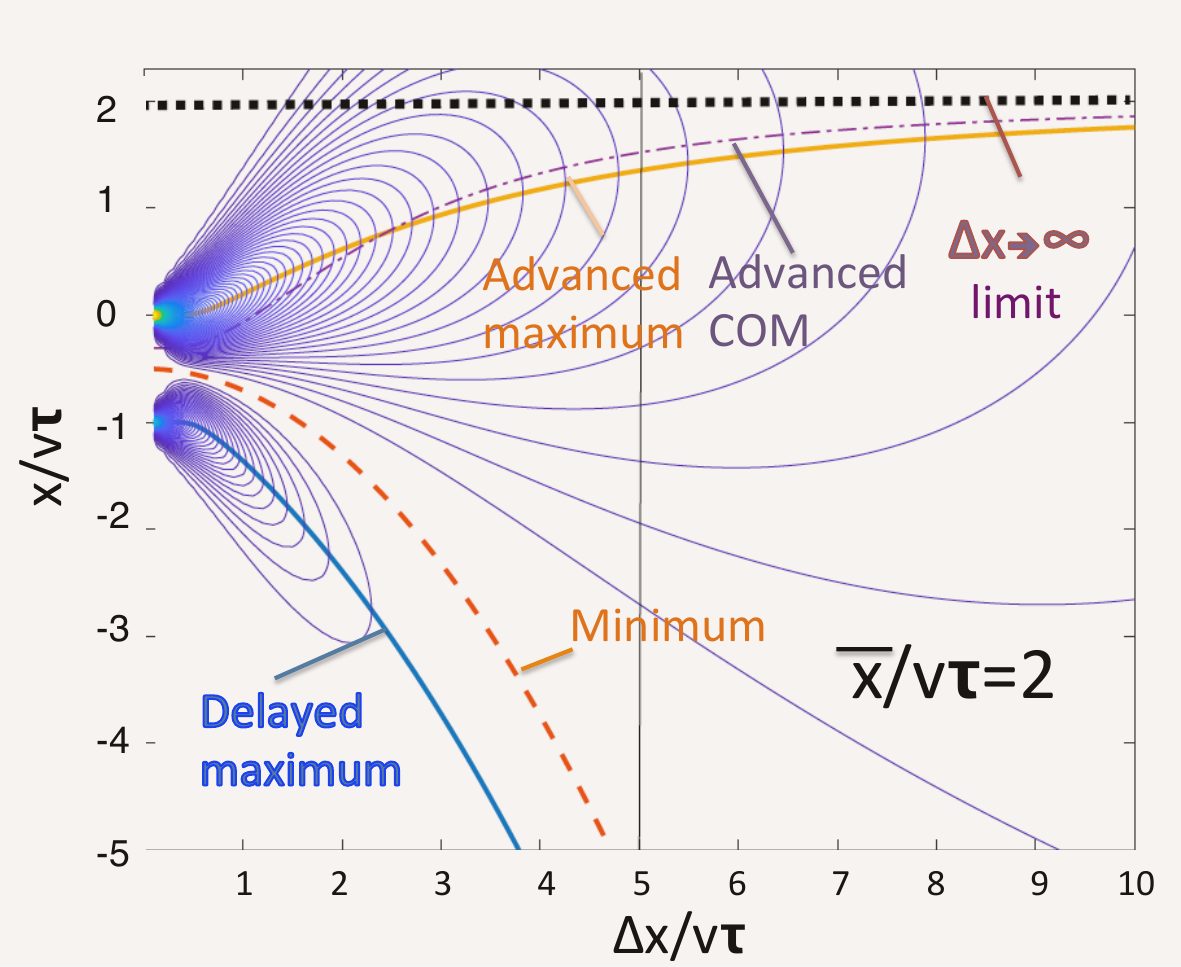}
\caption {Contour plot of the probability density $P(x)=|\A_1G(x)+\A_2G(x+v\t)|^2$ vs. $x$ and $\D x$ in the case where the second WP is delayed. The 
path amplitudes are chosen {\r so that $\A_1/\A_2=-1.5$}.
%$\A_1=0.5883$ and $\A_2=-0.3922$.
As the width $\D x$ increases, the  maximum and the minimum, located at $x=-v\t $ and $x=-v|\t|/2$ for $\D x << v\t$,  disappear.
%, accompanied by the minimum, located for $\D x << v\t$ at $x=-v|\t|/2$.
The surviving maximum is pushed forward, and asymptotically approaches $\overline x=2$ [cf. (\ref{4a})], indicated by the dotted line. 
Also shown is the centre of mass (COM) of the density, whose position coincides with that of the density's peak only in the limit $\D x \to \infty$. 
}
\end{figure}
 \begin{figure}[h]
\includegraphics[angle=0,width=8.5cm, height= 6cm]{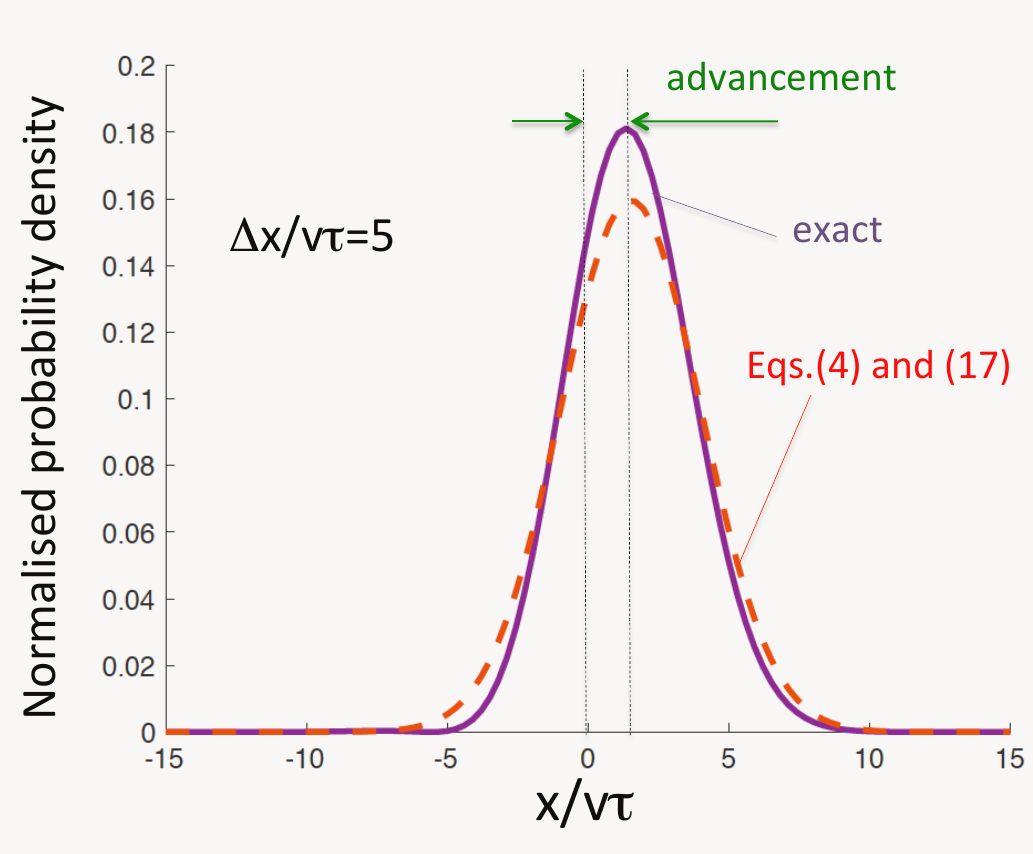}
\caption {Comparison between the normalised to unity density $|G_1(x)|^2$ in Eq.(\ref{2a})
and that obtained from  Eq.(\ref{3a}) with $\overline x$ given by Eq.(\ref{AA2}) for $\Delta x/v\t=5$.
The amplitudes $\A_{1,2}$ are the same as in Fig.2, and the advancement of the peak of about 
$1.35v\t$ is somewhat smaller than $2v\t$, achieved for $\Delta x/v\t\to \infty$.
}
\end{figure}
\newline
Furthermore, for $\overline x > 0$ the second of Eqs.(\ref{b5}) yields 
 \begin{eqnarray}\label{c1}
\frac{\A_2}{\A_1}= -\frac{\overline x}{\overline x+v\t}<0, 
\end{eqnarray}
so one converts a delay into advancement by means of destructive interference. 
In particular, as $\overline x \to \infty$, $\A_2\to -\A_1$, which makes the Gaussian  (\ref{3a}) very small, and detection by $D_1$ highly unlikely. 
\newline
%Importantly this is a case of \e{reshaping}. 
For the  ratio between $|G(x)|^2$ and that in Eq.(\ref{3a}) one finds
 \begin{eqnarray}\label{c2}
\frac{P(x)}{|G(x)|^2} = |\A_1|^2
%\times \q\q\q\q\q\q\q\q\q\n
\left |1-\frac{\overline x}{\overline x+v\t}\frac {G(x+v\t)}{G(x)}\right |^2 
\end{eqnarray}
where we have used (\ref{c1}).
For any $x>0$ and $\overline x>0$ the ratio (\ref{c2}) is less than unity.
In other words, the density (\ref{3a}), if advanced, will always fit under the front tail of 
the density, transmitted to $D_1$ through the left arm only (see Fig.4). 
\begin{figure}[h]
\includegraphics[angle=0,width=8.5cm, height= 6cm]{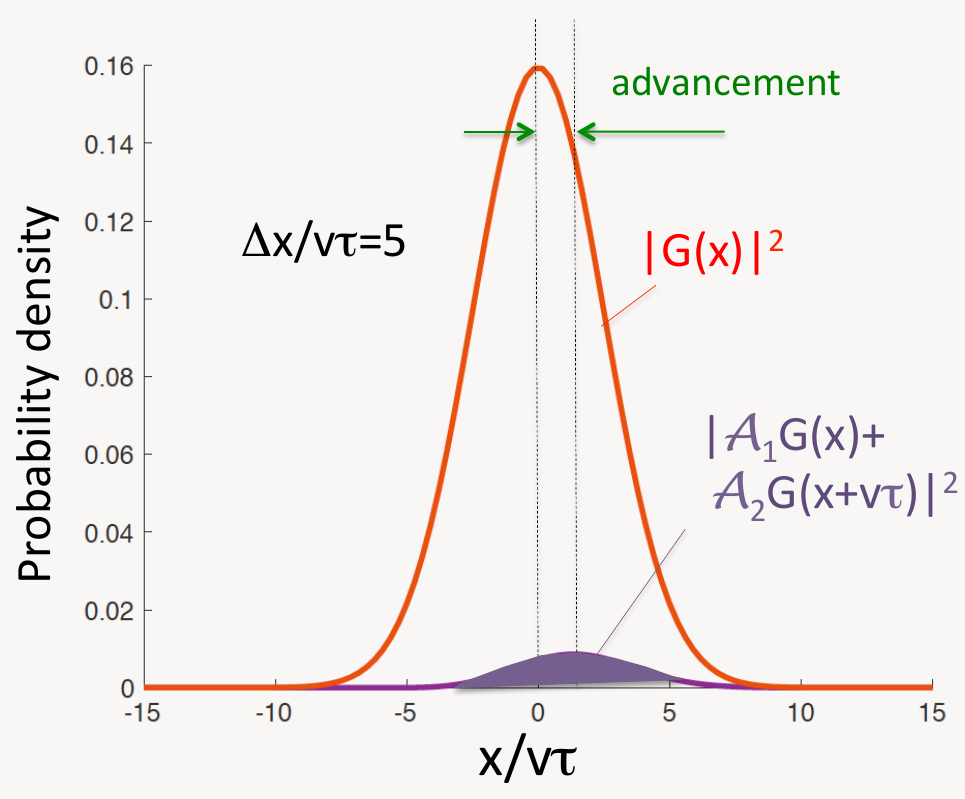}
\caption {A comparison between the  density $|G_1(x)|$ in Eq.(\ref{2a})
and 
the one obtained for the initial WP travelling along the left arm only. 
The parameters are the same as in Figs. 2 and 3.}
\end{figure}
\newline
Thus, for $\Delta x >>vt$ most particles are directed towards the detector $D_2$,  where  interference between the paths  is constructive.  The probability density is still given 
by Eqs.(\ref{3a})-(\ref{4a}),  with the amplitudes $\A_1$ and $\A_2$ replaced with $\A_3$ and $\A_4$, respectively.  The peak of the WP lies a distance
at $vt/2$ behind the peak propagating  through the left arm only, and the probability of detection by $D_2$ tends to $1$.
%%%%%%%%%%%%%%%%%%%%%%%
\section{Delaying the peak}
One can also prepare a setup where the transmitted peak lags behind the peak of the WP travelling through the left arm, $\overline x < 0$ , by choosing
 \begin{eqnarray}\label{d1}
\frac{\A_2}{\A_1}= -\frac{|\overline x|}{|\overline x|-v\t}.
\end{eqnarray}
The ratio is positive, and interference is constructive, for as long as the peak still lies ahead of the density $|G(x+v\t)|^2$, corresponding to the right arm only, 
 i.e., for $|\overline x| < v\t$. If one tries to delay by more than $v\t$ interference becomes destructive again. 
It is easy to see that the peak, delayed in this manner, fits under the rear tail of the delayed Gaussian , since
 \begin{eqnarray}\label{a4}
\frac{P(x)}{|G(x+v\t)|^2} = |\A_2|^2
%\times \q\q\q\q\q\q\q\q\q\n
\left |1-\frac{|\overline x|-v\t}{|\overline x|}\frac {G(x)}{G(x+vt)} \right |^2 <1\q\q
\end{eqnarray}
for all $x<-v\t$.
As with the advancement, for a large delay, $|\overline x|>> vt $, one has $\A_2\to -\A_1$, and detection by $D_1$ becomes highly improbable. 
%%%%%%%%%%%%%%%%%
\section{ \e{Superluminality},   \e{negative times} and \e{abnormal delays}}
So far there was nothing unusual in our analysis of the MZI interference. In neither case the particle was found in a place where it would not arrive if the 
same initial state were propagated along only one arm of the interferometer. 
 \newline
Spurious sensationalism will enter the discussion if, following \cite{WP0}, one decides to convert spatial 
displacements into temporal durations. [But, might object a fastidious reader, surely the small peak, advanced by $\overline x$, 
would arrive at a fixed detector $\overline x/v$ seconds earlier than the peak of the WP travelling along the left arm only?
Is it not an observable time interval?
Yes, but one cannot deduce from this temporal delay the duration the particle spends inside the MZI.]
 %this is not what we have in mind. The problems start if one tries to deduce from this temporal  delay, 
%    \begin{eqnarray}\label{e1}
%\t_{delay}= \overline x/v,
%\end{eqnarray}
% the duration the particle spends the MZI.
\newline 
 Here is a kind of reasoning which, in our opinion, one should try to avoid. 
Let the distance between the beamsplitters measured along the left arm of the MZI be $L$.
Then the maximum of the wave packet, travelling only along the left arm, will cover it in $L/v$ seconds.
If the peak (that is the maximum) of the wave packet transmitted across both arms and exiting along the destructive interference port is found
%If  the peak (that is the maximum), transmitted across both arms is found 
a distance $\overline x$ ahead, 
the particle \e{must have spent a shorter time between the beamsplitters}, 
    \begin{eqnarray}\label{e2}
\t_{inside}=L/v- \overline x/v. 
\end{eqnarray}
This we insist, is not a valid conclusion. 
One can arrange the delay $\t$,  $\A_1$, and $\A_2$, in order to ensure
% that the transmitted peak leaves
 that the transmitted peak, resulting from the interference between
 %Êinterference between 
 the forward tails of the Gaussians passing through the left and right arms,
  would 
  leave the MZI just as the maximum of the initial Gaussian enters it,
  %the MZI just as incident peak enters it,
   $ \overline x= L$. 
Then one may be tempted to claim that the MZI has been crossed infinitely fast, in defiance of Einstein's relativity. 
This, we argue, is not true, since if we block the right arm of the MZI, a wave packet traveling through the left arm only
at a perfectly subluminal speed $v<<c$
would also provide for the detection of particles at 
%the position where the peak of the interfering wave packet appears, 
the same place,
and would do so with a higher probability [cf. Eq.(\ref{c2})].
%This, we argue, is not true, since the particle could arrive at the same location, if the same
%while propagating along the left arm of the MZI at a perfectly subluminal speed $v$. 
\newline
Worse still, arranging for $ \overline x>L$ would make the duration $\t_{inside}$ negative,
so the transmitted peak leaves the MZI before the incident one enters  it.
Such a behaviour was reported, for example in \cite{BEF}.
Again, the same argument suggests that nothing untoward happens also in this case.
Besides, the notion of a \e{negative duration} is not defined, just as the concept 
of \e{negative probability}, which cannot be related to frequencies \cite{FeynNEG}.
% Rather than claiming \e{experimental evidence 
%of quantum negative time}, one should perhaps question the method leading to this 
%bizarre notion. 
\newline
Finally, the same applies to abnormally long delays, $\overline x << -v\t$, 
$\t_{inside} >>L/v$. One would also detect such particles in the tail of 
a broad wave packet, propagating only via the right arm of the MZI, 
where it experiences a much shorter delay $\tau$. 
%%%%%%%%%%%%%%%%%%%%%%%%%%%
\section{Conclusions}
It is probably fair to say that few authors, with possible exception of Nimtz and Stahlhofen \cite{NS}, 
are prepared to accept that classically forbidden quantum events, such as tunnelling, fall outside the remit 
of special relativity. More often the apparent speed up of the transmitted particle is related to some form 
of reshaping, which destroys the causal relationship  between the incoming and transmitted peaks of the
wave packet involved \cite{REV1},\cite{BW}. Indeed, the tunneling dynamics in a relativistic quantum field theory framework also displays a maximum of the tunneled wavepacket advanced (relative to the freely propagating maximum) 
while fully respecting relativistic causality \cite{qft1,qft2}. Naive, and purely classical examples of reshaping include  the train analogies, 
found in \cite{WIN}, and the paper-and-scissors metaphor of \cite{DSE}. 
\newline
A more realistic mechanism, considered here, relies on interference between the parts of a broad wave packet
travelling through different arms of a two-way interferometer. By arranging the path amplitudes, one can, in principle, 
place the  transmitted peak anywhere before, or behind, the peak of the WP travelling through one arm only. 
With a delay of $\t$ secs experienced in one arm, overall advancement of the transmitted peak requires the interference
to be destructive.
%between the parts of the . 
When recombined, the parts of the WP have opposite signs, and  cancel each other, leaving only a small  advanced peak.
{\r Note that experimentally, what is required is the detection of spatial shifts of the maximum of the 
	transmitted wavepacket \cite{sahoo}.}

It is a peculiar property of the Gaussians that a {\it  broad} advanced WP, although greatly reduced, repeats
the shape of the incident pulse.  The same mechanism, we note, is responsible for delaying the transmitted WP by more 
that $\t$. 
\newline
Spurious controversies may arise if one relies on  the position of the transmitted peak hoping to deduce the duration spent 
by the particle inside the interferometer.  It is, however, obvious that  there is neither  \e{superluminal behaviour}, or 
the possibility of \e{time travel} where $\t_{inside}$ in Eq.(\ref{e2}) turns negative. The simple reason is that
the particle is perfectly able to arrive at the same time in the same place in a perfectly  pedestrian manner, i.e.,  if the same initial state propagates  through  just one arm, at a purely subliminal velocity $v$. 
It is comforting, although not necessary for our argument, to note that the corresponding probability is reduced, whenever both arms of the interferometer are engaged, 
and the rest of the particles are re-directed towards the other detector $D_2$. 
\newline 
 We note further that at the centre of the controversy is quantum interference or, more precisely, the Uncertainty Principle.
 The Principle forbids revealing  the 
 path travelled by a particle without destroying interference between the paths in the process \cite{FeynL}. 
Similarly, with the interference intact, it does not allow one to combine the durations spent in each arm 
 into a single meaningful duration, describing the passage when both arm are available to the particle.  
 \newline
This is, in a nutshell, the tunnelling time problem. It is easy to show \cite{DSE}, \cite{NOT} that in 
the case of a potential barrier the transmitted state is given by a similar superposition of 
various copies of the freely propagating wave packet, subjected to  various spatial delays.
{\r Such a situation could potentially be simulated by a cascade of nested MZIs \cite{cascade}.}
Because of destructive interference, the tunnelling amplitude is small, although individual terms
in the superposition are not. The two-way analogue studied here captures the main features of the phenomenon, and we expect  
 most of the above conclusions to apply also in the case of tunnelling.
 It is difficult to imagine, given the simplicity of our example, how a further explanation can be given to such  a basic 
 destructive interference phenomenon.  
 %Anyone who succeeds in this task will not only solve the tunnelling time problem, 
 %but also resolve the double-slit conundrum, and unveil the \e{only mystery of quantum mechanics}, which fascinated Feynman in his time \cite{FeynL}.
 %%%%%%%%%%%%
\section{Appendix A}
The probability density in the port $D_1$ of the MZI in Fig.1a, 
and the net probability of detection by $D_1$
 are given by
\begin{eqnarray}\label{AA1}
P(x)=|\A_1|^2 |G(x)|^2+ |\A_2|^2 |G(x+v\t)|^2+\q\q\q\n
 2\R[\A_2^*\A_1]G(x)G(x+v\t),\q\q\q\q\q\q\q\n
P=|\A_1|^2 + |\A_2|^2 |+ 2\R[\A_2^*\A_1]\exp\left (-\frac{v^2t^2}{2\Delta x^2}\right ),\q\q
%\q\q\q\q
\end{eqnarray}
respectively.
For the mean $\overline x=\int xP(x)dx/P$ we have
\begin{eqnarray}\label{AA2}
\overline x=-v\t \frac{|\A_2|^2+\R[\A_2^*\A_1]\exp\left (-\frac{v^2t^2}{2\Delta x^2}\right )}{|\A_1|^2 + |\A_2|^2 |+ 2\R[\A_2^*\A_1]\exp\left (-\frac{v^2t^2}{2\Delta x^2}\right )}\q
\end{eqnarray}
As the ratio $v\t/\Delta x$ increases, $P(x)$ and $\overline x$ vary between $|\A_1|^2 + |\A_2|^2$ and $|\A_1+ \A_2|^2$, and 
$-v\t|\A_1|^2/(|\A_1|^2 + |\A_2|^2)$ and $-v\t \R[A_2/(A_1+\A_2)]$ [cf. Eq.(\ref{4a})], respectively.
%%%%%%%%%%%%%%%%%%%%%%%%%%%%%%%
\section*{Data availability}
\noindent
The datasets used and/or analysed during the current study available from the corresponding author on reasonable request.
 %%%%%%%%%%%%%%%%%%%
 \def\bibsection{\section*{\refname}}

%%%%%%%%%%%%%%%%%%%%%%%%%%%%%%%
\section*{Funding}
\noindent
DS acknowledges financial support by the Grant PID2021-126273NB-I00 funded by MICINN/AEI/10.13039/501100011033 and by \e{ERDF A way of making Europe}, as well as by the Basque Government Grant No. IT1470-22.
%%%%%%%%%%%%%%%%%%%%%%%%%%%%%%%
\section*{Author contributions}
\noindent
D. Sokolovski and A. Matzkin both prepared and reviewed the manuscript.
%%%%%%%%%%%%%%%%%%%%%%%%%%%%%%%
\section*{Competing interests}
\noindent
The authors declare no competing interests.
\end{document}